\documentclass[a4paper, oneside, twocolumn, notitlepage, 10pt]{extarticle_ecoc}
\usepackage{ecoc}
\usepackage{color,soul}
\usepackage{microtype}
\usepackage{epstopdf}
\usepackage{pgfplots}
\usepackage{csvsimple}
\usepackage{tikz}
\usepackage{tkz-fct}
\usepackage{pgfplots}
\usepgfplotslibrary{groupplots}
\usetikzlibrary{matrix,positioning}
\usepackage{tikz-dimline}
\usetikzlibrary{arrows.meta,decorations.pathreplacing}
\usetikzlibrary{shapes.misc}
\usepackage{import}

\pgfplotsset{compat=newest}
\pgfplotsset{
	grid style = {
    	dash pattern = on 0.025mm off 0.95mm on 0.025mm off 0mm,
    	line cap = round,
    	black,
    	line width = 0.5pt,
    	opacity = 0.4
  }
}

\pgfplotsset{
	every axis/.append style={
		axis lines* = left,
    	axis line style = {-{Triangle[length = 1.3mm, width = 1.3mm]}, line width = 0.265mm}  
	}
}

\usetikzlibrary{decorations.pathmorphing} 
\usetikzlibrary{matrix} 
\usetikzlibrary{calc} 

\tikzstyle{block} = [draw,rectangle,thick,minimum height=2em,minimum width=2em,align=center]
\tikzstyle{multi} = [circle,draw,cross]
\tikzstyle{connector} = [-Triangle,thick]
\tikzstyle{line} = [thick]
\tikzstyle{branch} = [coordinate]
\tikzstyle{guide} = []

\tikzset{%
  do path picture/.style={%
    path picture={%
      \pgfpointdiff{\pgfpointanchor{path picture bounding box}{south west}}%
        {\pgfpointanchor{path picture bounding box}{north east}}%
      \pgfgetlastxy\x\y%
      \tikzset{x=\x/2,y=\y/2}%
      #1
    }
  },
  cross/.style={do path picture={    
    \draw [line cap=round] (-1,-1) -- (1,1) (-1,1) -- (1,-1);
  }},
  plus/.style={do path picture={    
    \draw [line cap=round] (-3/4,0) -- (3/4,0) (0,-3/4) -- (0,3/4);
  }}
}

\begin{document}
\selectlanguage{english}    


\title{ On a Scalable Path for Multimode MIMO-DSP }%


\author{
    Fabio A. Barbosa\textsuperscript{(1)}, Stefan Rothe \textsuperscript{(1,2)},
    Dennis Pohle\textsuperscript{(2)},
    J\"urgen W. Czarske\textsuperscript{(2)}, Filipe M. Ferreira\textsuperscript{(1)}
}

\maketitle                  


\begin{strip}
 \begin{author_descr}

   \textsuperscript{(1)} Optical Networks, Dept Electronic \& Electrical Eng, University College London, UK,
   \textcolor{blue}{\uline{f.ferreira@ucl.ac.uk}}

  \textsuperscript{(2)} Laboratory of Measurement and Sensor System Technique, Faculty of Electrical and Computer Engineering, TU Dresden, 01062 Dresden, Germany


 \end{author_descr}
\end{strip}

\setstretch{1.1}
\renewcommand\footnotemark{}
\renewcommand\footnoterule{}


\begin{strip}
  \begin{ecoc_abstract}
    
A novel MIMO-DSP for space-division multiplexing over multimode fibres is proposed. 
A principal modes approach is shown to provide two-fold benefits: over 13 times channel memory reduction while minimising the number of optical front-ends needed to detect a subset of the spatial domain.
    \textcopyright2022 The Authors
    

  \end{ecoc_abstract}
\end{strip}


\section{Introduction}
Space-division multiplexing (SDM) has emerged as a solution to overcome the capacity limit of single-mode fibres (SMFs) \cite{Richardson_2013}. Amongst SDM approaches, multi-mode fibres (MMFs) offer the highest spatial information density and thus potential integration gains in transceivers, optical amplifiers, WSSs, and both fibre- and chip-to-fibre interface. 
One of the main challenges with multimode SDM is that conventional MIMO equalization requires all guided modes to be detected for successful equalisation \cite{Winzer:outage} -- binding the number of fibre modes with the number of coherent front-ends required at transceivers. 
This prevents the installation of many-mode ($\gg1$) fibres since it would not be possible to deploy transceivers with as many optical front-ends from day one. 
And given that new fibre deployments need a major motivation, this tie needs to be breakdown. Motivation such as that offered by MMFs and multiplexers approaching 1000 spatial and polarisation modes \cite{fil:ECOC1000, Fontaine:21}.

\begin{figure*}[h!]
    \centering
    \def\svgwidth{\linewidth}
    \fontsize{8pt}{7.2}\selectfont
    \resizebox{0.98\textwidth}{!}{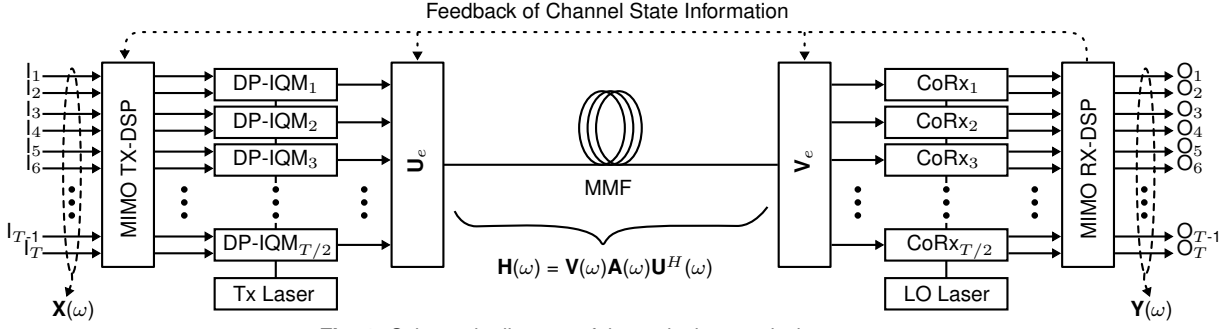}
    \caption{Schematic diagram of the optical transmission system.}
    \label{fig:TransmissionSystem}
\end{figure*}

In our project \cite{ucl_2021}, we aim to develop a new family of scalable transceivers to exploit a growing number of modes over the same fibre infrastructure over several transceiver generations. We are investigating programmable solutions for an optical mapping of spatially independent signals to a desired combination of modes. This process can be performed using a spatial light modulator (SLM) and multi-plane light conversion (MPLC) \cite{Fontaine:prog}. We explore different approaches for designing  mode multiplexers including both the established wavefront matching approach and an intelligent camera-based solution with actor neural networks~\cite{pohle2022}.

Here, we present a novel MIMO-DSP strategy capable of transmitting $T$ spatial and pol. tributaries over a $M$-mode fibre, with $M$ several times larger than $T$ -- a principal modes (PMs) inspired approach is demonstrated with the benefit of reducing the length of the channel impulse response. 



\section{Principal Modes}


The MMF channel in Fig.~\ref{fig:TransmissionSystem} can be described by a frequency-dependent $M \times M$ matrix $\textbf{H}(\omega)$, where $M$ is the number of spatial and pol. modes. $\textbf{H}(\omega)$ can account for effects of transceivers and other components of the transmission link. Using $\textbf{H}(\omega)$ a group delay operator can be defined as \cite{Askarov:15}
\begin{equation}\label{eq:GD}
\textbf{G}(\omega) = j\frac{\partial \textbf{H}(\omega)}{\partial \omega}\textbf{H}^{-1}(\omega).
\end{equation}
The eigenvectors and eigenvalues of $\textbf{G}(\omega)$ correspond to the \textit{input} PMs and their group delays, respectively. The \textit{output} PMs can be determined from $\textbf{H}(\omega)$ and the input PMs. Since by definition PMs are  frequency independent to 1st order, over a certain frequency interval, i.e. a coherence bandwidth, we can assume that \cite{Fan:05}
\begin{equation}\label{eq:ChannelDecomposition}
    \textbf{H}(\omega) = \textbf{V}\boldsymbol\Lambda(\omega)\textbf{U}^{H},
\end{equation}
where $\textbf{U}$ and $\textbf{V}$ are the input and output PMs, respectively, $(\cdot)^{H}$ is the Hermitian and $\boldsymbol\Lambda(\omega)$ is a diagonal matrix accounting for the overall group delays and mode gain/loss. 
The latter is given by the logarithm of the eigenvalues of $\textbf{D}(\omega) = \textbf{H}(\omega)\textbf{H}^{H}(\omega)$.

Here, we work under the assumption that channel information can be fed back from the receiver to the transmitter, as depicted in Fig.~\ref{fig:TransmissionSystem}, allowing the SDM system to take advantage of the PMs. 
In such favourable conditions, our objective is to assess the possibility to develop a MIMO-DSP approach capable of reducing the channel memory and suppressing the modal crosstalk at the front of the receiver. An effort focused at reducing to $T$ the number of optical front-ends necessary to transmit/detect $T$ spatial tributaries (with $T$ several times smaller than $M$-modes). 
Outside of this work is the problem of tracking of a dynamic channel (with realistic latency) and the respective impact on the performance of a transmission system as in Fig.~\ref{fig:TransmissionSystem}. In any case, our results show progress in dealing with the limit bandwidth of PMs by sub-band processing  tributaries.


\section{Estimation of  the PMs}
\label{estPMs}


To estimate the MIMO channel $\textbf{H}(\omega)$ for subsequent calculation of the PMs, we periodically insert training sequences (TSs) into the data transmitted in each tributary. Mutually orthogonal TSs are chosen so that training overhead can be minimised.
The $L$-length TS on the $p$-th tributary is given by the inverse discrete Fourier transform (IDFT) of a frequency-domain sequence defined as
\begin{equation}\label{eq:TSs}
    S_{p}[k] = 
    \begin{cases}
        C[k], & k = p + lM \\
        0,    & otherwise,
    \end{cases}
\end{equation}
where $k$ is the discrete frequency index, $l = [0,1,\cdots,\lfloor L/M-1 \rfloor]$. And $C[k]$, in this work, is a binary phase shift keying (BPSK) sequence. From (\ref{eq:TSs}), it follows that each tributary uses a different set of discrete frequencies to transmit the TSs. 
The time-domain TSs are then padded on each side by prefixes. At the receiver, a least-square (LS) frequency domain channel estimation is performed. The channel between the $j$-th transmitter and the $i$-th receiver, i.e. $\hat{H}_{i,j}[k]$, is calculated as
    $\hat{H}_{i,j}[k] = R_i[k]/S_j[k], \: k = j + lN,$
where $R_i[k]$ is the TS received at the $i$-th receiver. 
An additional interpolation and extrapolation step is required to estimate the channel over all frequency positions. Similar strategies can be exploited for OFDM signals. Averaging over multiple TSs, we reduce the impact of additive white Gaussian noise (AWGN) on the estimates of $\textbf{H}(\omega)$. Then, we estimate the input and output PMs through (\ref{eq:GD}) and (\ref{eq:ChannelDecomposition}).




The estimation of the PMs can be severely affected by artefacts on the channel estimation. In this work, we also modify the output PMs by zero-forcing the residual channel, i.e. the end-to-end channel obtained after using the PMs for the transmission. This new set of PMs (including both input and output PMs) is indicated hereafter by PMs*. In addition, we also consider slicing the signal bandwidth in frequency blocks and estimate a different set of PMs for the central frequency of each block.


\section{Results and Discussion}

We performed transmission simulations over a 50-km MMF (with 12 spatial and polarisation modes) using the parameters in Table (\ref{tab:MMFParam}) -- using the semi-analytical channel model in \cite{fil:semiJLT}. The integrated modal crosstalk was assumed to be -20 dB/km. 

Input and output PMs (or PMs*), $\textbf{U}_e$ and $\textbf{V}_e$, respectively, are estimated at the receiver. At the transmitter, tributaries are optically multiplexed to the MMF using $\textbf{U}_e$ assuming an ideal feedback path. 
At the receiver, tributaries are optically de-multiplexed with $\textbf{V}_e$ before residual channel estimation with the strategy described above. Note that, $\textbf{U}_e$ and $\textbf{V}_e$ are applied in the optical domain assuming an ideal programmable mode multiplexer \cite{Fontaine:prog}.
Finally, we performed minimum mean-squared error (MMSE) MIMO equalisation to the received signals. We transmit 16-QAM signals at 33 GBd and use TSs with 4096 symbols. 

\begin{table}[h]
\centering
\caption{MMF parameters per LP mode.} \label{tab:MMFParam}
\begin{tabular}{|c|c|c|c|}
\hline
\small
LP Mode & $\alpha$ {[}$\frac{dB}{km}${]} & DMD {[}$\frac{ps}{km}${]} & CD {[}$\frac{ps}{nm \times km}${]} \\ \hline
01   & 0.1913               & 0              & 22.1761                      \\ \hline
02   & 0.1747               & -10.5358       & 21.5473                      \\ \hline
11a  & 0.1830               & -1.4619        & 22.1516                      \\ \hline
11b  & 0.1830               & -1.4619        & 22.1516                      \\ \hline
21a  & 0.1747               & 10.2386        & 21.8434                      \\ \hline
21b  & 0.1747               & 10.2686        & 21.8434                      \\ \hline
\end{tabular}
\end{table}

\begin{figure}[t]
    \centering
    \small
    \begin{tikzpicture}
        \definecolor{color1}{RGB}{0,113,188}
        \definecolor{color2}{RGB}{216,82,24}
        \definecolor{color3}{RGB}{0,0,0}
        \definecolor{color4}{RGB}{0,127,0}
        \definecolor{color5}{RGB}{126,47,142}
        \definecolor{color6}{RGB}{255,0,0}
        \begin{axis}[
            width  = \linewidth,
            height = 6cm,
            ymode = log,
            xlabel = {Sample index},
            ylabel = {CIR (norm.)},
            xmin   = -20,
            xmax   = 20,
            ymax   = 1,
            grid    = both,
            scaled x ticks=false,
            tick label style={/pgf/number format/fixed},
            legend style={nodes={scale=0.9, transform shape},line width=0.4mm,at={(0.2,0.99)},anchor=north,draw=none,fill=none,text width=5em},
            legend columns = 1,
            legend cell align={left}]
            
            \addlegendentry{$h[n]$}
            \addlegendentry{PMs}
            \addlegendentry{PMs*}
            
            
            \addplot[solid, mark options={solid}, color3, line width = 0.4mm] table[x=xAxis,y=sum_H,col sep=comma]{ChannelImpulseResponse.csv};
            
    
            \addplot[solid, mark = o, mark options={solid}, color1, line width = 0.4mm] table[x=xAxis,y=sum_VHU,col sep=comma]{ChannelImpulseResponse.csv};
            
            
            \addplot[dashed, mark options={solid}, color2, line width = 0.4mm] table[x=xAxis,y=sum_VnewHU,col sep=comma]{ChannelImpulseResponse.csv};
            
            
            

    	\end{axis}
    \end{tikzpicture}
    \vspace{-0.3cm}
    \caption{CIR of the actual channel (solid line) and of the residual channel when PMs (solid line with circles) and PMs* (dashed line) are used. Response integrated over all paths.}
    \label{fig:CIR}
\end{figure}
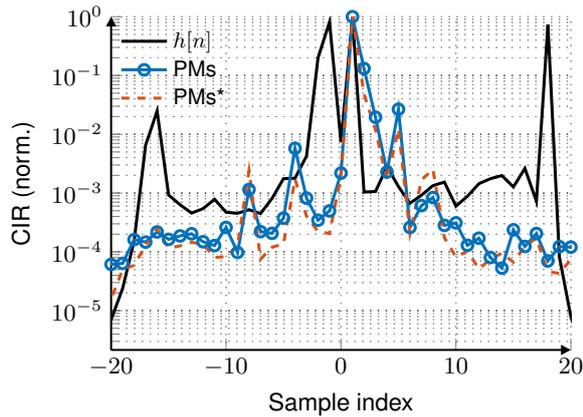

Fig. \ref{fig:CIR} shows the channel intensity impulse response (CIR) at the receiver input for: just MMF $h[n]$ (solid line), with PMs (line with circles), and with PMs* (dashed line). Note that one single set of PMs (or PMs*) is used for the entire signal bandwidth. The results show a significant compression of the CIR when using PMs (or PMs*).

To investigate the transmission of $T$ tributaries over a $M$-mode fibre with $M>T$ but using only $T$ optical-front ends, we simulated the transmission of 6 tributaries using the 6 best PMs (lowest group delays) over Tab.~\ref{tab:MMFParam} MMF. Note that, $T$ and $M$ refer to spatial and pol. modes -- thus for dual pol. front-ends, only $T/2$ are required/used.
At the receiver, the residual channel is estimated using only the 6 detected tributaries to perform MMSE MIMO equalisation. 
Performance is evaluated considering the received constellations SNR ($E{|X|^2}/E{|X-Y|^2}$).

\begin{figure}[t]
    \centering
    \small
    \begin{tikzpicture}
        \definecolor{color1}{RGB}{0,113,188}
        \definecolor{color2}{RGB}{216,82,24}
        \definecolor{color3}{RGB}{0,0,0}
        \definecolor{color4}{RGB}{0,127,0}
        \definecolor{color5}{RGB}{126,47,142}
        \definecolor{color6}{RGB}{255,0,0}
        \begin{axis}[
            width  = \linewidth,
            height = 6.7cm,
            xlabel = {OSNR [dB]},
            ylabel = {SNR [dB]},
            xmin   = 25,
            xmax   = 45,
            ymin   = 15,
            ymax   = 40.8,
            grid    = both,
            scaled x ticks=false,
            tick label style={/pgf/number format/fixed},
            legend style={nodes={scale=0.9, transform shape},line width=0.4mm,at={(0.5,1.1)},anchor=north,draw=none,fill=none,text width=5em},
            legend columns = -1,
            legend cell align={left}]
            
            
            
            \addplot[solid, mark options={solid}, color3, line width = 0.4mm] table[x=OSNR_dB,y=SNRThdB,col sep=comma]{Results_33GBd_OSNR_45dB_to_25dB_1Blocks_6Trib.csv};
            
            
            \addplot[dash dot, mark options={solid}, color4, line width = 0.4mm] table[x=OSNR_dB,y=SNRValueSVD,col sep=comma]{Results_33GBd_OSNR_45dB_to_25dB_1Blocks_6Trib.csv};
            
    
            \addplot[solid, mark = triangle, mark options={solid}, color1, line width = 0.4mm] table[x=OSNR_dB,y=SNRValueVHUEq,col sep=comma]{Results_33GBd_OSNR_45dB_to_25dB_1Blocks_6Trib.csv};
            
            
            \addplot[solid, mark = o, mark options={solid}, color2, line width = 0.4mm] table[x=OSNR_dB,y=SNRValueVnewHUEq,col sep=comma]{Results_33GBd_OSNR_45dB_to_25dB_1Blocks_6Trib.csv};
            
    
            \addplot[dashed, mark = triangle, mark options={solid}, color1, line width = 0.4mm] table[x=OSNR_dB,y=SNRValueVHUEq,col sep=comma]{Results_33GBd_OSNR_45dB_to_25dB_1Blocks_12Trib.csv};
            
    
             \addplot[dashed, mark = o, mark options={solid}, color2, line width = 0.4mm] table[x=OSNR_dB,y=SNRValueVnewHUEq,col sep=comma]{Results_33GBd_OSNR_45dB_to_25dB_1Blocks_12Trib.csv};

            
            
            
            \node[text width = 6cm,align = center, font=\small,above,rotate = 33] at (axis cs:35,30.8) {Theory};
            
            \node[text width = 6cm,align = center, font=\small] at (42,32) {PMs*};
            
            \node[text width = 6cm,align = center, font=\small] at (42,26) {PMs};
            
            \node[text width = 6cm,align = center, font=\small] at (42,35) {SMs};
            

    	\end{axis}
    \end{tikzpicture}
    \caption{Received constellations SNR as a function of OSNR when SMs (dash-dotted line), PMs (lines with triangles) and PMs* (lines with circles) are used. Solid lines with markers for the 6 PMs case, while the dashed lines for the 12 PMs case.}
    \label{fig:SNRxOSNR}
\end{figure}
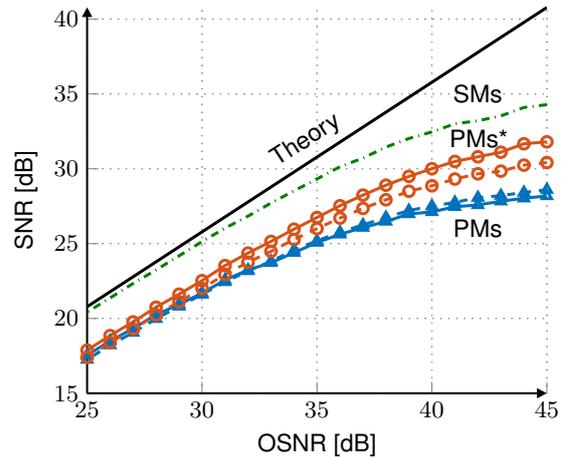

Results are shown in Fig.~\ref{fig:SNRxOSNR} for 6 and 12 tributaries transmission, both cases with MMSE for the residual channel.
For comparison, it is also shown the theoretical SNR and the SNR using the Schmidt modes (SMs) for the transmission of 12 tributaries -- SMs are obtained through the singular value decomposition (SVD) of the estimated $\textbf{H}(\omega)$ \cite{kahn:MIMO}. 
Performance in both scenarios considered, i.e. 6 and 12 tributaries, is similar, and both PMs and PMs* performance is close to ideal SMs.
Note that, the results in Fig.~\ref{fig:SNRxOSNR} (as in Fig.~\ref{fig:CIR}) were obtained with a single set of PMs along the entire signal bandwidth, making full use of PMs bandwidth. 
Since a single vector mode mapping is necessary, this strategy is  implementable in the optical domain, e.g using a SLM and MPLC at both transmission ends. 
Consequently, to transmit $T$ spatial tributaries one could excite $T$ input PMs and use $T$ (or $T/2$ dual pol.) receivers to detect their corresponding output PMs. 
Validating the potential for PMs to be exploited as a platform for flexible scale of multimode SDM transceivers. 

Although SMs outperform PMs (and PMs*), they are not readily suitable for optical domain implementation due to their frequency-dependency. And, for a digital domain implementation of the SMs, suppression of modal crosstalk or  reduction of channel memory would not be achieved at the receiver front-end. 

\begin{figure}[h]
    \centering
    \small
    \begin{tikzpicture}
        \definecolor{color1}{RGB}{0,113,188}
        \definecolor{color2}{RGB}{216,82,24}
        \definecolor{color3}{RGB}{0,0,0}
        \definecolor{color4}{RGB}{0,127,0}
        \definecolor{color5}{RGB}{126,47,142}
        \definecolor{color6}{RGB}{255,0,0}
        \begin{axis}[
            width  = \linewidth,
            height = 6cm,
            xlabel = {Num. of Blocks},
            ylabel = {CIR Compression},
            xmin   = 1,
            xmax   = 17,
            ymin   = 2,
            ymax   = 14,
            xtick = {1,5,...,15},
            ytick = {2,5,...,15},
            grid    = both,
            scaled x ticks=false,
            tick label style={/pgf/number format/fixed},
            legend style={nodes={scale=0.9, transform shape},line width=0.4mm,at={(0.5,1.1)},anchor=north,draw=none,fill=none,text width=5em},
            legend columns = -1,
            legend cell align={left}]
            
            
            
            \addplot[solid, mark = triangle , mark options={solid}, color1, line width = 0.4mm] table[x=Blocks,y=CIRCompVHU,col sep=comma]{CIRCompression.csv};
            
            
            \addplot[dashed, mark = o , mark options={solid}, color2, line width = 0.4mm] table[x=Blocks,y=CIRCompVnewHU,col sep=comma]{CIRCompression.csv};
            
            
            \node[text width = 6cm,align = center, font=\small] at (15,9.1) {PMs*};
            
            \node[text width = 6cm,align = center, font=\small] at (15,13) {PMs};
            

    	\end{axis}
    \end{tikzpicture}
    \caption{CIR compression vs the number of frequency blocks when PMs (solid line) and PMs* (dashed line) are used.}
    \label{fig:CIRCompression}
\end{figure}
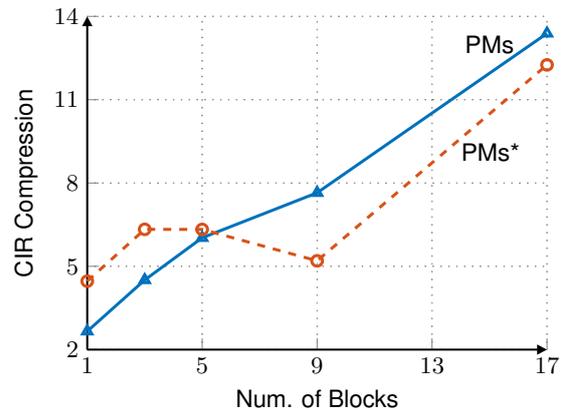

Finally, we evaluate the case for slicing the signal bandwidth and using different sets of PMs for each block, aiming at further CIR compression which in turn would allow for higher DMD fibres and/or lower equalisation complexity \cite{Beril:Complexity}. 
Fig. \ref{fig:CIRCompression} shows an increasing CIR compression vs \#blocks, reaching a compression over 13 times.
Although challenging, for a limited number of frequency bins, programmable multiplexers might be scaled up \cite{Mounaix_2020}.  

\section{Conclusions}
A novel MIMO-DSP inspired on PMs has been demonstrated to reduce channel memory over 13x while using only $T$ spatial and pol. tributaries and $T$ (or $T/2$ dual pol.) optical-front ends at the receiver for a $M$-mode fibre (with $M$ several times $T$). This approach opens the way to a new transceivers capable scaling the number of spatial tributaries in line with ever-growing traffic demand.


\section{Acknowledgements}
This work was supported by the UKRI Future Leaders Fellowship MR/T041218/1 and the German Research Foundation with grant no. (CZ~55/42-2). Underlying data at doi.org/10.5522/04/19739365.


\printbibliography

@article{Richardson_2013,
	doi = {10.1038/nphoton.2013.94},
	year = 2013,
	month = {apr},
	volume = {7},
	number = {5},
	pages = {354--362},
	author = {D. J. Richardson and J. M. Fini and L. E. Nelson},
	title = {Space-division multiplexing in optical fibres},
	journal = {Nature Photonics}
}

@INPROCEEDINGS{Fontaine:prog,
  author={Fontaine, Nicolas K. and Chen, Haoshuo and Ryf, Roland and Neilson, David and Alvarado, Juan Carlos and van Weerdenburg, John and Amezcua-Correa, Rodrigo and Okonkwo, Chigo and Carpenter, Joel},
  booktitle={2017 European Conference on Optical Communication (ECOC)}, 
  title={Programmable Vector Mode Multiplexer}, 
  year={2017},
  volume={},
  number={},
  pages={1-3},
  doi={10.1109/ECOC.2017.8346099}}

@inproceedings{Fontaine:21,
author = {Nicolas K. Fontaine and Haoshuo Chen and Mikael Mazur and Lauren Dallachiesa and K.W. Kim and Roland Ryf and David Neilson and Joel Carpenter},
booktitle = {Optical Fiber Communication Conference (OFC) 2021},
journal = {Optical Fiber Communication Conference (OFC) 2021},
pages = {M3D.4},
publisher = {Optica Publishing Group},
title = {Hermite-Gaussian mode multiplexer supporting 1035 modes},
year = {2021},
url = {http://opg.optica.org/abstract.cfm?URI=OFC-2021-M3D.4},
doi = {10.1364/OFC.2021.M3D.4},
}

@article{Fan:05,
author = {Shanhui Fan and Joseph M. Kahn},
journal = {Opt. Lett.},
number = {2},
pages = {135--137},
publisher = {OSA},
title = {Principal modes in multimode waveguides},
volume = {30},
month = {Jan},
year = {2005},
url = {http://opg.optica.org/ol/abstract.cfm?URI=ol-30-2-135},
doi = {10.1364/OL.30.000135},
}

@INPROCEEDINGS{fil:ECOC1000,
  author={Ferreira, Filipe M. and Barbosa, Fabio A.},
  booktitle={European Conference on Optical Communication }, 
  title={Towards 1000-mode Optical Fibres}, 
  year={2022},
  volume={},
  number={},
  pages={},
  doi={submitted}}

@ARTICLE{fil:semiJLT,
  author={Ferreira, Filipe Marques and Costa, Christian S. and Sygletos, Stylianos and Ellis, Andrew D.},
  journal={Journal of Lightwave Technology}, 
  title={Semi-Analytical Modelling of Linear Mode Coupling in Few-Mode Fibers}, 
  year={2017},
  volume={35},
  number={18},
  pages={4011-4022},
  doi={10.1109/JLT.2017.2727441}}

@ARTICLE{kahn:MIMO,
  author={Ho, Keang-Po and Kahn, Joseph M.},
  journal={Journal of Lightwave Technology}, 
  title={Linear Propagation Effects in Mode-Division Multiplexing Systems}, 
  year={2014},
  volume={32},
  number={4},
  pages={614-628},
  doi={10.1109/JLT.2013.2283797}}

@article{Beril:Complexity,
author = {Beril Inan and Bernhard Spinnler and Filipe Ferreira and Dirk van den Borne and Adriana Lobato and Susmita Adhikari and Vincent A. J. M. Sleiffer and Maxim Kuschnerov and Norbert Hanik and Sander L. Jansen},
journal = {Opt. Express},
number = {10},
pages = {10859--10869},
publisher = {OSA},
title = {DSP complexity of mode-division multiplexed receivers},
volume = {20},
month = {May},
year = {2012},
url = {http://opg.optica.org/oe/abstract.cfm?URI=oe-20-10-10859},
doi = {10.1364/OE.20.010859},
}

@article{Winzer:outage,
author = {Peter J. Winzer and Gerard J. Foschini},
journal = {Opt. Express},
number = {17},
pages = {16680--16696},
publisher = {OSA},
title = {MIMO capacities and outage probabilities in spatially multiplexed optical transport systems},
volume = {19},
month = {Aug},
year = {2011},
url = {http://opg.optica.org/oe/abstract.cfm?URI=oe-19-17-16680},
doi = {10.1364/OE.19.016680},
}

@misc{ucl_2021, title={Beyond Exabit Optical Communications: From new devices, via New Dimensions to new systems}, url={http://www.ucl.ac.uk/iccs/beyond-exabit}, journal={Institute of Communications and Connected Systems}, author={Ucl}, year={2021}, month={Jun}}

@inproceedings{pohle2022,
  title={Intelligent Self Calibration Tool for Adaptive Mode Multiplexers using Multiplane Light Conversion},
  author={Pohle, Dennis and Rothe, Stefan and Barbosa, Fabio and M. Ferreira, Filipe and Koukourakis, Nektarios and Czarske, Jürgen W.},
  booktitle={25th Congress of the International Commission for Optics},
  year={2022},
  organization={International Society for Optics and Photonics},
  note={submitted}
}

@article{Askarov:15,
author = {Daulet Askarov and Joseph M. Kahn},
journal = {J. Lightwave Technol.},
number = {19},
pages = {4032--4038},
publisher = {OSA},
title = {Long-Period Fiber Gratings for Mode Coupling in Mode-Division-Multiplexing Systems},
volume = {33},
month = {Oct},
year = {2015},
url = {http://opg.optica.org/jlt/abstract.cfm?URI=jlt-33-19-4032},
}

@article{Mounaix_2020,
	doi = {10.1038/s41467-020-19601-3},
	url = {https://doi.org/10.1038/s41467-020-19601-3},
	year = 2020,
	month = {nov},
	publisher = {Springer Science and Business Media {LLC}},
	volume = {11},
	number = {1},
	author = {Mickael Mounaix and Nicolas K. Fontaine and David T. Neilson and Roland Ryf and Haoshuo Chen and Juan Carlos Alvarado-Zacarias and Joel Carpenter},
	title = {Time reversed optical waves by arbitrary vector spatiotemporal field generation},
	journal = {Nature Communications}
}

\vspace{-4mm}

\end{document}